\documentclass[a4paper]{jpconf}
\usepackage{graphicx}
\usepackage{epstopdf}
\usepackage{color}

\newcommand{\msol}{\,$M_{\odot}$}

\begin{document}
\title{Probing the parameter space of HD~49933: a comparison between global and 
local methods}

\author{O L Creevey$^{1,2}$, M Bazot$^{3,4}$} 
\address{$^1$ Instituto de Astrof\'isica de Canarias (IAC), E-38200 La Laguna, Tenerife, Spain}
\address{$^2$ Dept. de Astrof\'isica, Universidad de La Laguna (ULL), E-38206 La Laguna, Tenerife, Spain}
\address{$^3$ Centro de Astrof\'isica da Universidade do Porto, Rua das Estrelas, 4150-762 Porto, Portugal}
\address{$^4$ Institut for Fysik og Astronomi, Aarhus Universitet, DK-8000 Aarhus C., Denmark}
\ead{orlagh@iac.es, bazot@astro.up.pt}




\begin{abstract}
We present two independent methods for studying the global stellar
parameter space (mass $M$, age, chemical composition $X_0$, $Z_0$) of HD~49933 with
seismic data. 
Using a local minimization and 
an MCMC algorithm, we obtain consistent results
for the determination of the stellar properties: M = 1.1--1.2 \msol, Age~$\sim$~3.0~Gyr, 
$Z_0\sim0.008$. 
A description of the
error ellipses can be defined using Singular Value Decomposition
techniques, and this is validated by comparing the errors with those
from the MCMC method.  

\end{abstract}

\section{Introduction}

HD~49933 is a main sequence solar-type star that was observed by CoRoT. It is the first stellar object where solar-like oscillations were clearly detected in the photometric signal \cite{app08,gru09}. 
There has been some controversy over the original labelling of the oscillation frequencies with their
mode degrees, and in this work we present results based on the data published by \cite{app08},
although we note that since this publication there has been a preference among the scientific community towards an inverted
mode-labelling \cite{gru09}.  
We present two independent methods of fitting the observational data, while placing an emphasis on defining the boundaries of the parameter space where the model of this star lies.
The best-fitting models are determined using the Levenberg-Marquardt (LM) and Markov Chain Monte Carlo (MCMC) algorithms, while 
the uncertainties and the form of the parameter space that is constrained by the set of observations is described by both
Singular Value Decomposition (SVD) and MCMC. 

We use the frequencies from \cite{app08}, and we compute the frequency separations to use 
as the observational data to model.  
The stellar models are computed using ASTEC \cite{jcd08a} which include the EFF equation of state, OPAL 95 opacities together with Kurucz low-T opacities, NACRE reaction rates and overshooting, and the oscillation frequencies are calculated with ADIPLS \cite{jcd08b}.




\section{Local minimization of stellar observables}

\begin{figure}[h!]
\begin{center}
\includegraphics[width=0.43\textwidth]{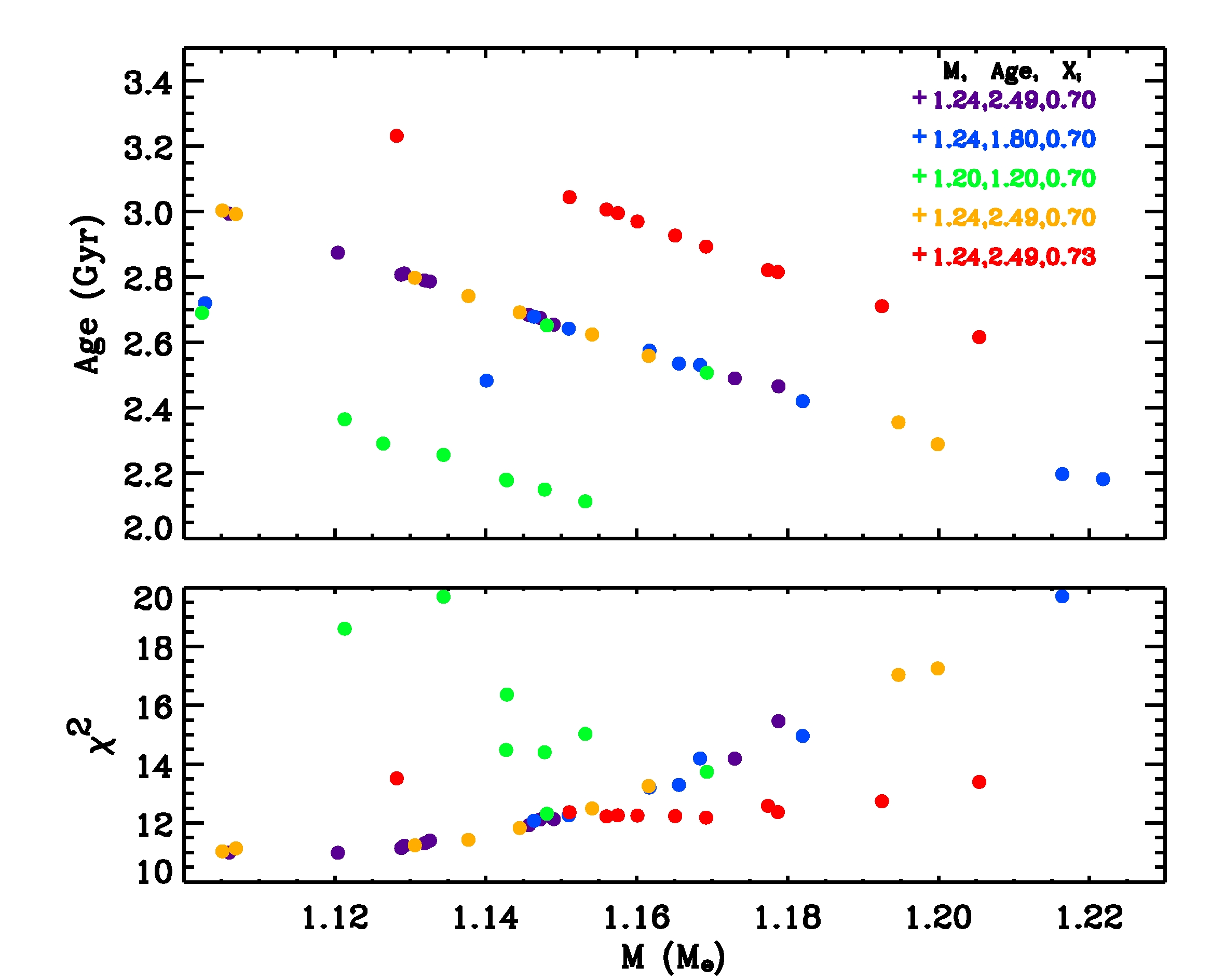}
\includegraphics[width=0.43\textwidth]{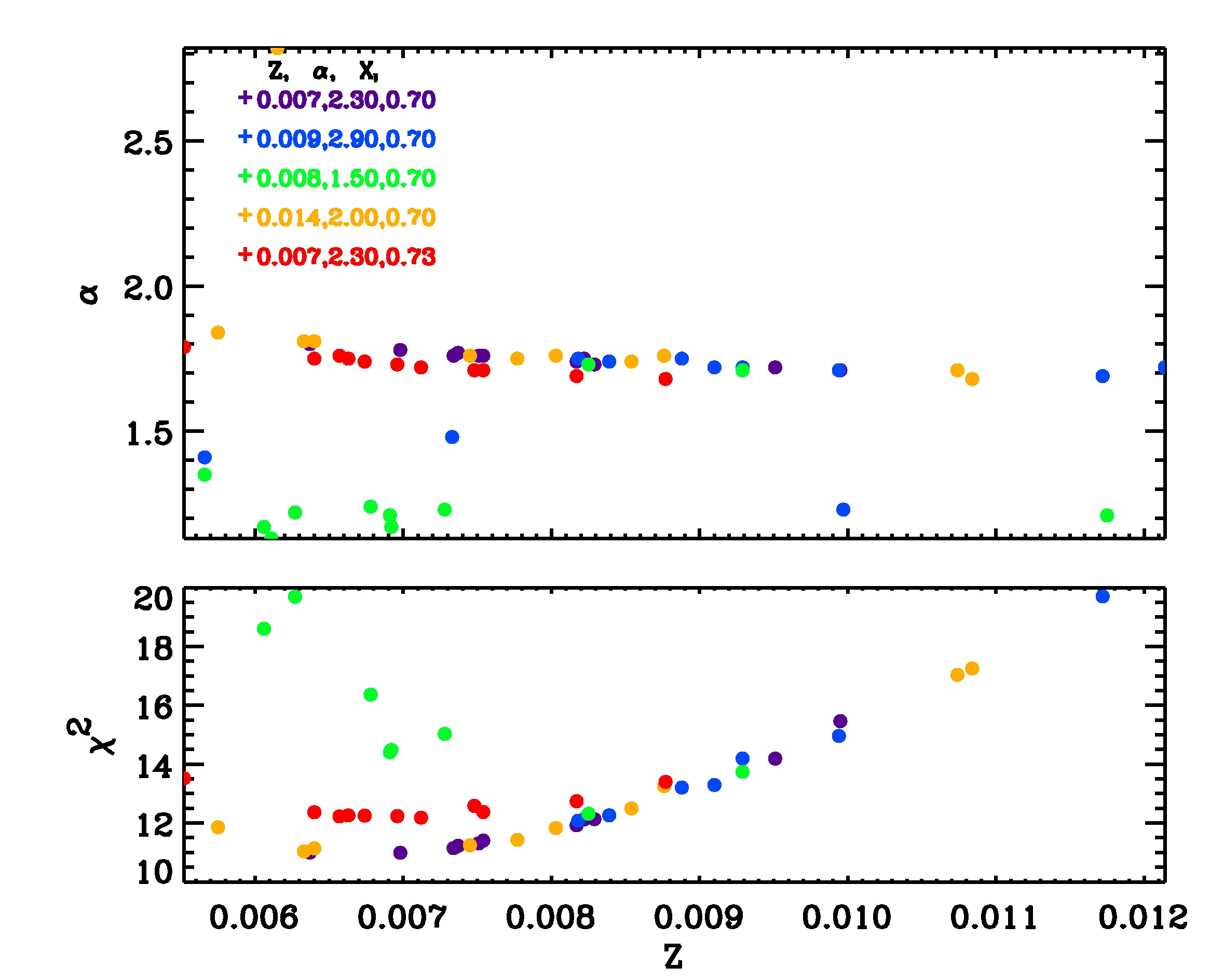}
\end{center}
\caption{Levenberg-Marquardt results for the best-fitting parameters using the $\ell = 0$ large frequency separations and $T_{\rm eff}, L_{\odot}, [M/H]$ data from \cite{app08} as constraints on the models.  The various colours indicate different initial guesses of the parameters, all having fixed $X_0 = 0.70$ except for the red dots which have fixed $X_0 = 0.73$. Each colour has a set of solutions, because we also varied the initial guess of the mass. \label{fig:lm}
}
\end{figure}

Using as observational constraints the $\ell = 0$ frequency separations, 
$T_{\rm eff}, L_{\odot},[M/H]$ from \cite{app08}, we obtain the following set
of stellar parameters for HD~49933 using LM:
$M = 1.12$ M$_{\odot}$, Age = 2.9 Gyr, $Z_0 = 0.007$, $\alpha = 1.80$, and $R =$ 1.39 R$_{\odot}$. 
$X_0 = 0.70$ was fixed, and the analysis was repeated using various
initial parameter values \cite{Creevey07}. 
In Fig.~\ref{fig:lm}, each dot is the result of a minimization, clearly showing a 
sensitivity to the initial guess.  This sensitivity is
not uncommon for local methods, however,
when we repeated our analysis using the data from \cite{gru09} we found a rather stable
solution of $M = 1.20$ M$_{\odot}$, Age = 2.5 Gyr, $Z = 0.0105$, $\alpha = 1.70$ and $R =$ 1.44 R$_{\odot}$.

\section{Global stellar parameter solution} 

Markov Chain Monte Carlo (MCMC) algorithms allows one to perform stochastic samplings of probability densities using properties of Markov Chains. This is a Bayesian methodology and we use it to estimate the posterior Probability Density Function (PDF) of the parameter(s) of our stellar model.  
The strength of this approach is that, once the PDF is derived, we can apply the classical tools of statistical inference to estimate the stellar parameters and associated confidence intervals --- a long-standing problem in stellar physics \cite{Jorgensen05,Takeda07,Creevey07}. 
This methodology was applied successfully to $\alpha$~Cen A \cite{Bazot08}.

We ran the MCMC algorithm using two different values
of initial parameters ($M = 1.1, 1.2$ \msol, $Z_0 = 0.007, 0.010$) while 
including a prior on $Z_0$ for the first run only.
Using the same observational constraints as those described in Sect.~2, we computed the marginal distributions
of the global stellar properties. 
Using their mean values as the best-fitting parameter and 
considering the standard deviation of the sample as the corresponding 1-$\sigma$ error bar, 
we obtain $M = 1.104 \pm 0.010$ M$_{\odot}$, Age = $3.16 \pm 0.11$Gyr, $Z = 0.0065 \pm 0.0006$, 
and $\alpha = 2.002 \pm 0.080$.  
 However, we also find that the second MCMC run hints at a possible (much less populated) alternative solution with a mass near 1.2 \msol.

\section{SVD describing the error ellipses}

When we obtain the best-fitting set of parameters $P$, then locally the models can be described as linear. With this assumption, we proceed to calculate the 
SVD  (${D = UWV^T}$) of the matrix 
 $D = \frac{\partial O}{\partial P}\sigma^{-1}$, where $O$ are the observables and $\sigma$ the
 observational errors.  The matrix $V$ describes the $N$-dimensional (in our case $N = 4)$ parameter correlations (essentially 4~4-element vectors), while $W$ describes the magnitudes of these vectors.
Fig.~\ref{fig:svd1} shows the (longest) two-dimensional projections of the four-dimensional vectors, for the parameters of $M$, age and $Z_0$. These 2-D vectors are represented by the red arrows in the figures, and the black ellipses are the error-ellipses defined by these vectors. 

\section{Uncertainties}
\begin{figure}[h!]
\begin{center}
\includegraphics[width=0.33\textwidth]{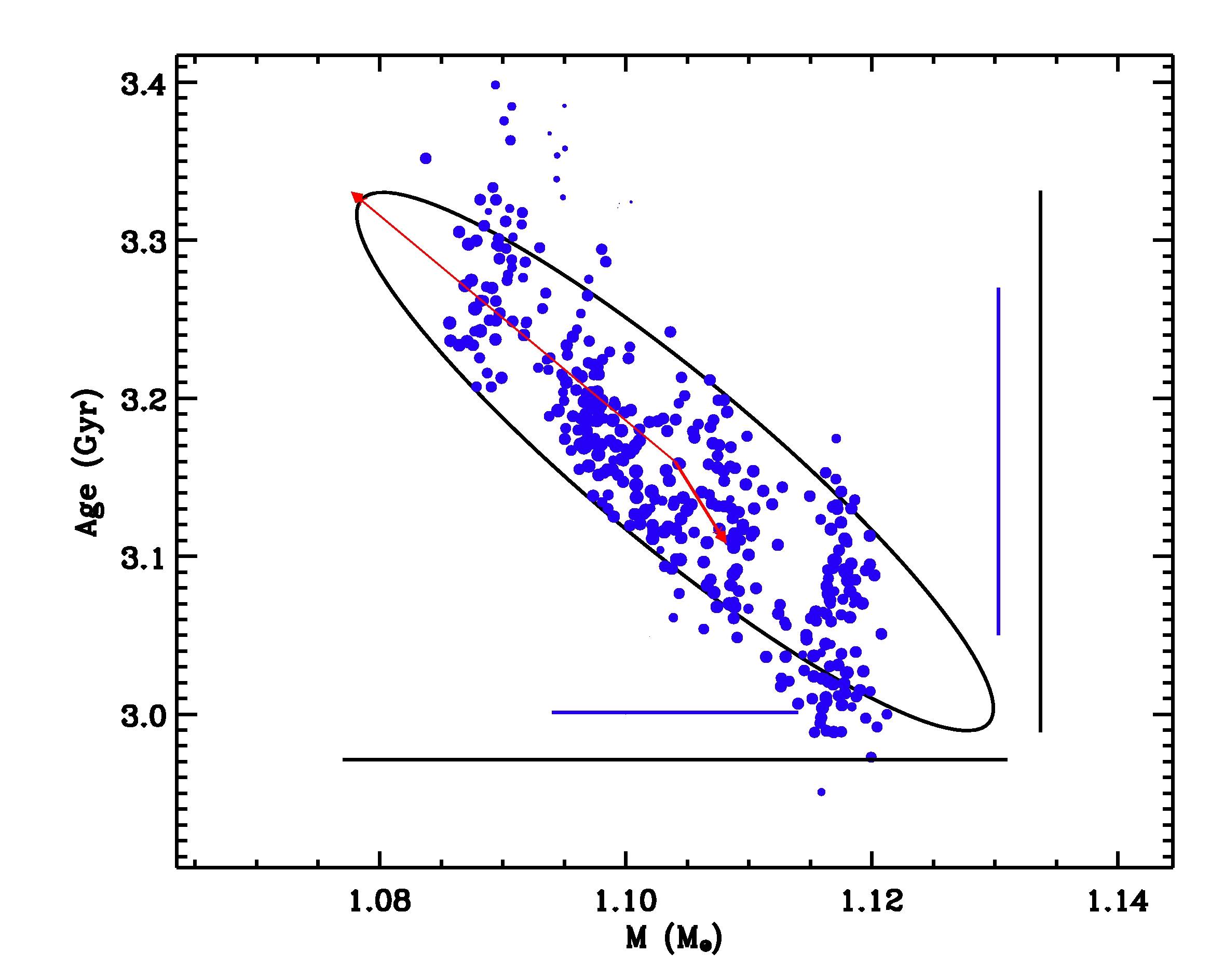}
\includegraphics[width=0.33\textwidth]{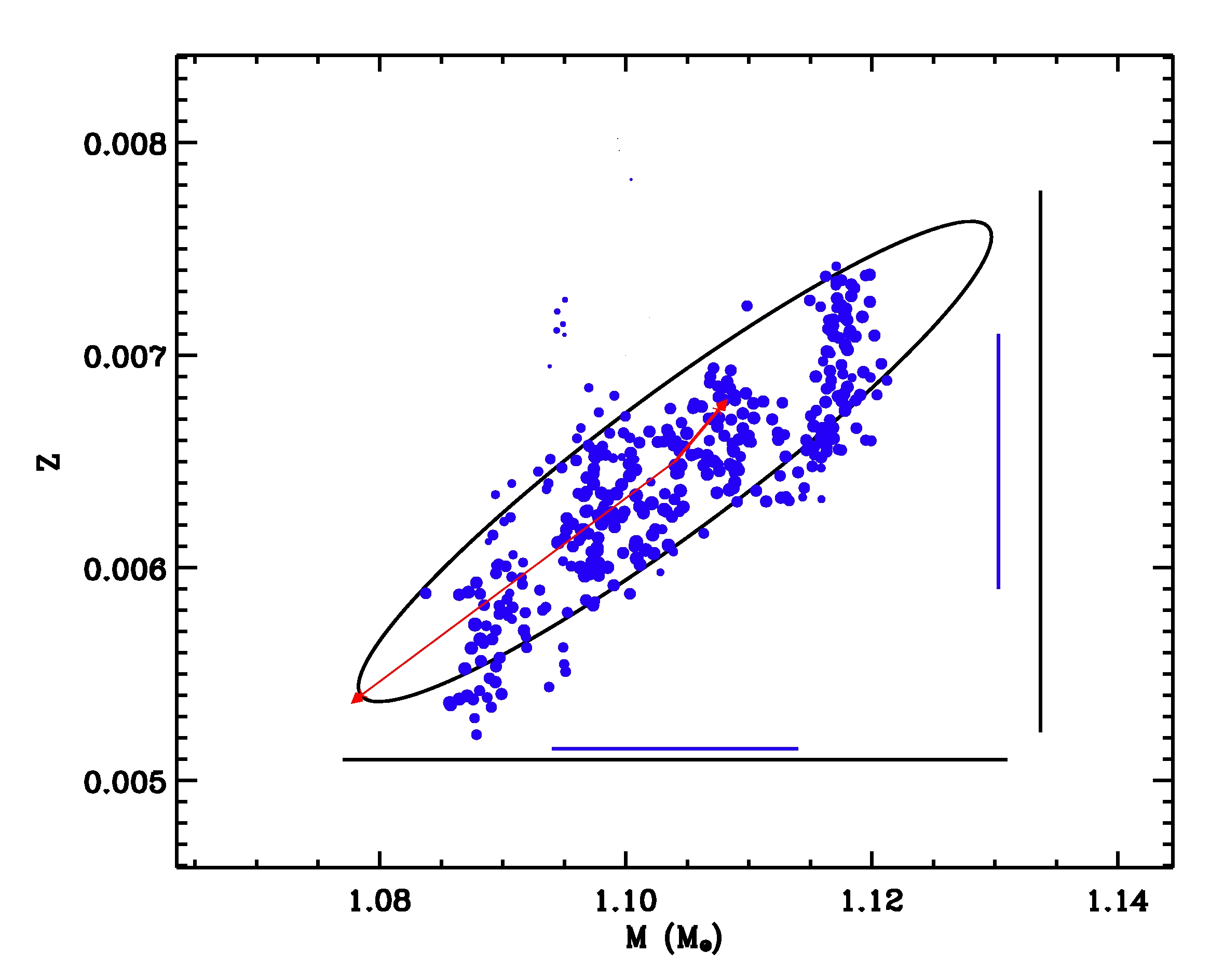}
\end{center}
\caption{2-D probability distributions for $M$ and Age (left) and $M$ and $Z_0$ (right) from the MCMC run.  
The size of the blue dots is proportional to the likelihood of the corresponding model. The red arrows are the two-dimensional scaled projections of $M$ and Age as defined by SVD, while the black ellipse is the 
analytical error-ellipse given by these vectors.\label{fig:svd1}}
\end{figure}

Fig.~\ref{fig:svd1} shows the error ellipses described by SVD.  The models generated from the MCMC runs are represented by the blue dots, with the size of the dots proportional to the likelihood values calculated from the MCMC (i.e. larger dots are more likely to be the true models). 
The error ellipses clearly encloses most of the (good) models from the MCMC method. This is strong evidence that, although we are assuming linearity for these models, the analytical approximation describes quite well the correlations that we expect to find, and the magnitudes of these correlations.  

\begin{figure}[h!]
\begin{center}
\includegraphics[width=0.28\textwidth]{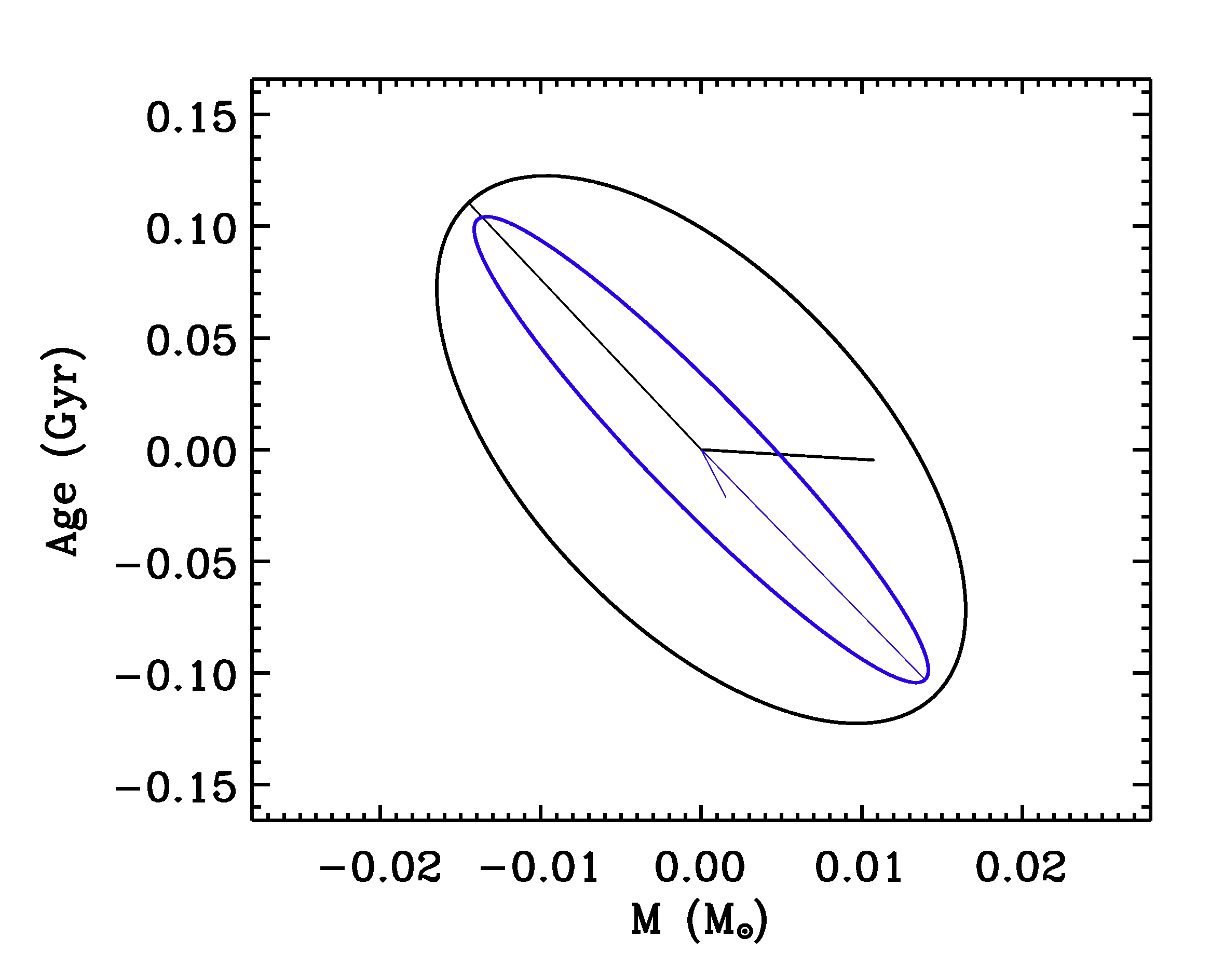}
\includegraphics[width=0.28\textwidth]{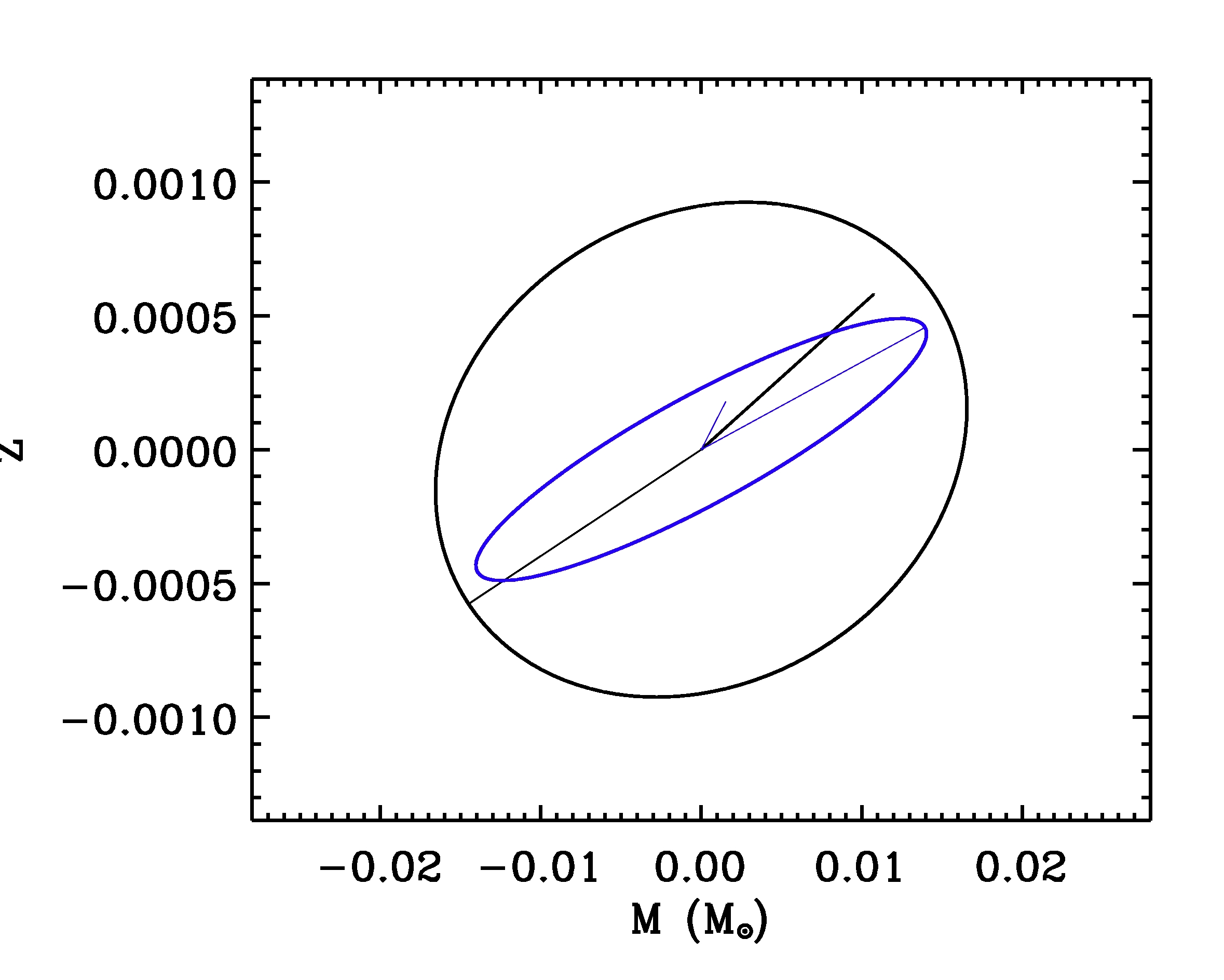}
\includegraphics[width=0.28\textwidth]{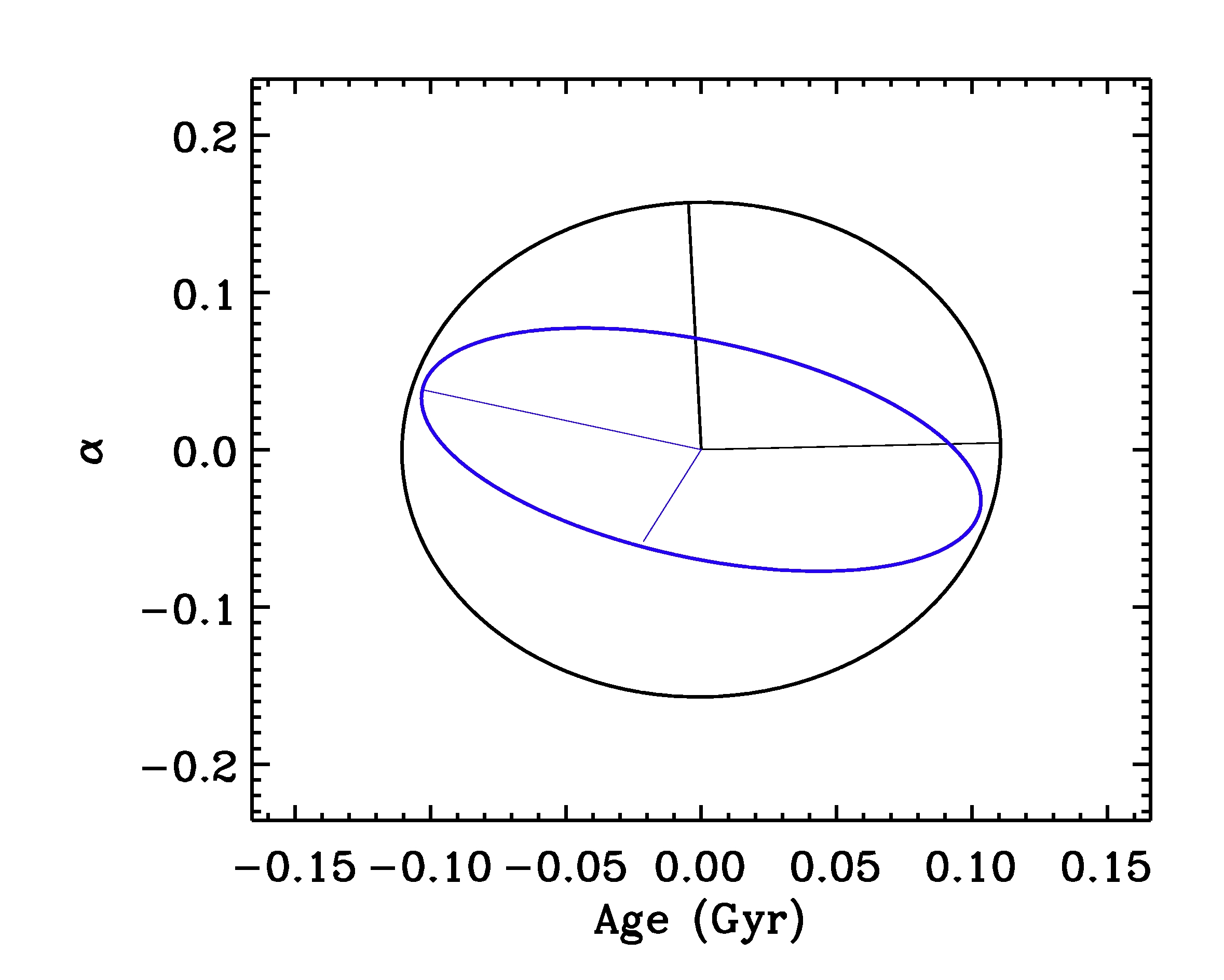}
\end{center}
\caption{Error ellipses for $M$ and Age (left), $M$ and $Z_0$ (centre) and Age and $\alpha$ (right) as defined by SVD, assuming a set of constraints comprising $T_{\rm eff}, L_{\odot}$ and the large frequency separations for $\ell = 0$ only (black), and then including the $\ell = 1$ frequency separations (blue).\label{fig:svd2}}
\end{figure}

Because SVD gives a good indication of the shape of the parameter spaces, we can use it to investigate the effects of including extra observational constraints.  In Fig.~\ref{fig:svd2} we compare the two-dimensional error-ellipses for various parameters when using the same constraints as Fig.~\ref{fig:svd1} (black ellipses), and then including the large frequency separations from the $\ell = 1$ modes (blue ellipses). The allowed parameter spaces defined by the extra set of constraints is reduced in volume by a factor of 2 for the 4 parameters, and the individual uncertainties are reduced by a significant amount for $Z_0$ and $\alpha$.


\pagebreak

\section{Conclusions}
We have analysed the seismic data from HD~49933 and we have found that the global stellar properties 
using two independent methods (local and global) are in agreement.  
Using the mode-identification and frequencies given by \cite{app08} and using the frequency
separations as the observational data to compare to the models we obtain 
$M = 1.12$ M$_{\odot} \pm 2\%$, Age = 2.9 Gyr $\pm$ 5\%, $X_0 = 0.70$ (fixed), 
$Z_0 = 0.007 \pm 18\%$, $\alpha = 1.80 \pm 9\%$, and 
$R =$ 1.39 R$_{\odot}$ from the local minimization method.  
This solution is supported by the results from the global method: 
$M=1.1$ \msol, Age = 3.16 Gyr, $Z_0$ = 0.0065 and $\alpha = 2.00$.
The MCMC run also finds a possible alternative solution with $M = 1.2$ \msol.
A similar solution is also 
found from the local method while using the reversed mode-tagging \cite{gru09}:
$M = 1.20$ M$_{\odot}$, Age = 2.5 Gyr, $Z = 0.0105$, $\alpha = 1.70$ and $R =$ 1.44 R$_{\odot}$.

We also studied the parameter space defining the correlations and uncertainties using the analytical
approach of SVD.
The results from the global MCMC validated the analytical description.
In particular we found that including the $\ell = 1$ frequency separations should reduce the 
volume of parameter space by at least a factor of two in all dimensions, with the individual parameter uncertainties reducing significantly for $Z_0$ and $\alpha$.




\ack
This research was in part supported by the European Helio- and Asteroseismology Network (HELAS), a major international collaboration funded by the European Commission's Sixth Framework Programme.

\end{document}